\documentclass[aps,prl,reprint,preprintnumbers,superscriptaddress,amsmath,amssymb,bibnotes,longbibliography]{revtex4-2}
\usepackage{graphicx}
\usepackage{dcolumn}
\usepackage{bm}
\usepackage{color}
\usepackage{booktabs}
\usepackage{ulem}
\usepackage{siunitx} 
\usepackage[colorlinks,linkcolor=blue,anchorcolor=blue,citecolor=blue,breaklinks,CJKbookmarks=True,urlcolor=blue,filecolor=blue,menucolor=blue,runcolor=blue]{hyperref}

\begin{document}

\title{Evolution of Magnetism in Ce$_{4}$Ge$_{7}$ under Magnetic Field and Pressure}

\author{Kaixin Ye}
\thanks{These authors contributed equally to this work.}
\affiliation  {New Cornerstone Science Laboratory, Center for Correlated Matter and School of Physics, Zhejiang University, Hangzhou 310058, China}

\author{Qihe Yu}
\thanks{These authors contributed equally to this work.}
\affiliation  {Hubei Key Laboratory of Photoelectric Materials and Devices, School of Materials Science and Engineering, Hubei Normal University, Huangshi 435002, China}

\author{Yongjian Li}
\affiliation  {New Cornerstone Science Laboratory, Center for Correlated Matter and School of Physics, Zhejiang University, Hangzhou 310058, China}

\author{Yanan Zhang}
\affiliation  {New Cornerstone Science Laboratory, Center for Correlated Matter and School of Physics, Zhejiang University, Hangzhou 310058, China}

\author{Ye Chen}
\affiliation  {New Cornerstone Science Laboratory, Center for Correlated Matter and School of Physics, Zhejiang University, Hangzhou 310058, China}

\author{Rui Li}
\affiliation  {New Cornerstone Science Laboratory, Center for Correlated Matter and School of Physics, Zhejiang University, Hangzhou 310058, China}

\author{Lin Jiao}
\affiliation  {New Cornerstone Science Laboratory, Center for Correlated Matter and School of Physics, Zhejiang University, Hangzhou 310058, China}

\author{M. Smidman}
\affiliation  {New Cornerstone Science Laboratory, Center for Correlated Matter and School of Physics, Zhejiang University, Hangzhou 310058, China}

\author{Yongjun Zhang}
\affiliation  {Hubei Key Laboratory of Photoelectric Materials and Devices, School of Materials Science and Engineering, Hubei Normal University, Huangshi 435002, China}

\author{Yu Liu}
\email[Corresponding author: ]{liuyuccm@zju.edu.cn}
\affiliation  {New Cornerstone Science Laboratory, Center for Correlated Matter and School of Physics, Zhejiang University, Hangzhou 310058, China}

\author{Huiqiu Yuan}
\email[Corresponding author: ]{hqyuan@zju.edu.cn}
\affiliation  {New Cornerstone Science Laboratory, Center for Correlated Matter and School of Physics, Zhejiang University, Hangzhou 310058, China}
\affiliation  {Institute of Fundamental and Transdisciplinary Research, Zhejiang University, Hangzhou 310058, China}
\affiliation  {Institute for Advanced Study in Physics, Zhejiang University, Hangzhou 310058, China}
\affiliation  {State Key Laboratory of Silicon and Advanced Semiconductor Materials, Zhejiang University, Hangzhou 310058, China}
\affiliation  {Collaborative Innovation Center of Advanced Microstructures, Nanjing 210093, China}

\date{\today}

\begin{abstract}

We report the magnetic and transport properties of single-crystalline Ce$_4$Ge$_7$, which crystallizes in a non-centrosymmetric orthorhombic structure (space group \textit{C222$_{1}$}). It undergoes an antiferromagnetic transition at $T_{\text{N}} = \SI{7.3}{K}$. When a magnetic field is applied along the \textbf{\textit{b}}-axis, a metamagnetic transition occurs at $\SI{1.3}{T}$ (at $\SI{2}{K}$). Correspondingly, this transition gives rise to an anomalous Hall effect, which is dominated by the intrinsic Karplus–Luttinger mechanism. Under hydrostatic pressure, $T_{\text{N}}$ of Ce$_4$Ge$_7$ initially increases slightly and is then gradually suppressed. 
For pressures above $P_{\text{c}} = \SI{10.2}{GPa}$, no magnetic order is observed.
The divergence of the $A$ coefficient, the maximum of the residual resistivity, and the non-Fermi liquid behavior around $P_{\text{c}}$ indicate the possible existence of an antiferromagnetic quantum critical point in Ce$_4$Ge$_7$.

\end{abstract}

\maketitle


\section{\uppercase\expandafter{\romannumeral1}. INTRODUCTION}

Ce-based heavy-fermion compounds have attracted considerable interest due to their rich magnetic properties, quantum critical behavior, and unconventional superconductivity \cite{wengMultipleQuantumPhase2016,siHeavyFermionsQuantum2010,pfleidererSuperconductingPhasesElectron2009}. The magnetism typically originates from the Ce 4$f$ electrons, and its ground state is determined by the competition between the Kondo interaction and the Ruderman-Kittel-Kasuya-Yosida (RKKY) interaction. When the Kondo interaction dominates, the localized 4$f$ electrons hybridize with itinerant conduction electrons, forming heavy quasiparticles with a large effective mass and screening the local moments, thereby favoring a paramagnetic ground state. Conversely, long-range magnetic order occurs when the RKKY interaction prevails. Due to the low energy scales of these interactions, the ground state can be effectively tuned through parameters such as chemical doping, dimensionality, magnetic field, or pressure, potentially driving the system toward novel quantum states such as quantum critical point, unconventional superconductivity, etc. \cite{gegenwartMagneticfieldInducedQuantum2002,lohneysenNonfermiliquidBehaviorHeavyfermion1994,chenHighpressureStudiesHeavy2016,wuDimensionalityTuningHeavyfermion2026}.
 
Binary CeGe$_{2-x}$ with a large Sommerfeld coefficient of $60\text{--}120~\mathrm{mJ/mol\cdot K^2}$ exhibits rich structural and magnetic properties \cite{matthiasSuperconductivityFerromagnetismIsomorphous1958,yashimaThermalMagneticProperties1982,moriNewDenseKondo1985,gokhaleCeGeCeriumGermaniumSystem1989,lambert-andronCrystalStructureProperties1990,schobinger-papamantellosStructuresMagneticProperties1991,lambert-andronCoexistenceOrderedDisordered1994,linThermalMagneticProperties2002,zanStudyMagneticOrdering2003,nakanoElectricalResistivitySpecific2005,zhangSynthesisStructuralCharacterization2013,budkoPhysicalPropertiesCeGe2x2014,jayasekaraComplexMagneticOrdering2014}. As summarized in Table~\ref{tab:CeGe_comparison}, when the Ge deficiency $x$ is less than 0.2, CeGe$_{2-x}$ crystallizes in the orthorhombic \textit{Imma} structure. For $x$ greater than 0.3, the compound adopts the tetragonal \textit{I4$_1$/amd} structure. Interestingly, the existence of superstructures was proposed in CeGe$_{2-x}$ \cite{yashimaThermalMagneticProperties1982,lambert-andronCrystalStructureProperties1990,lambert-andronCoexistenceOrderedDisordered1994,zhangSynthesisStructuralCharacterization2013}, which has not been experimentally confirmed, particularly for $x$ in the intermediate range of 0.2--0.3. In analogical lanthanide compounds $RE$Ge$_{2-x}$ with Ge deficiency of $x = 0.25$, such as Nd$_4$Ge$_7$ \cite{venturiniNewOrderedThSi2type1999}, Pr$_4$Ge$_7$ \cite{shcherbanCrystalStructureCompound2009}, and Sm$_4$Ge$_7$ \cite{zhangSynthesisStructuralCharacterization2013}, vacancy ordering leads to a superstructure (orthorhombic Nd$_4$Ge$_7$-type, space group \textit{C222}$_1$), expanding the unit cell to four times that of the parent phase. In addition, the magnetism of CeGe$_{2-x}$ was observed to be sensitive to the sample quality and stoichiometry.
For instance, polycrystalline CeGe$_{2}$ was first reported to be a ferromagnet with the Curie temperature ($T_{\text{C}}$) of $\SI{4.5}{K}$ \cite{matthiasSuperconductivityFerromagnetismIsomorphous1958}, while subsequently two magnetic transitions were proposed, an antiferromagnetic (AFM) one at $T_{\text{N}} = \SI{7}{K}$ and a ferromagnetic (FM) one at a lower $T_{\text{C}}$ of $\SI{4.3}{K}$ \cite{linThermalMagneticProperties2002}. Single-crystalline CeGe$_{1.75}$ was reported to be an antiferromagnet with $T_{\text{N}} = \SI{7}{K}$ \cite{zhangSynthesisStructuralCharacterization2013}. However, CeGe$_{1.76}$ with similar stoichiometry shows complex magnetic ordering \cite{budkoPhysicalPropertiesCeGe2x2014,jayasekaraComplexMagneticOrdering2014}, forming an incommensurate AFM below $T_{\text{N}} = \SI{7}{K}$, which changes to a commensurate AFM below $\SI{5}{K}$, with a different commensurate AFM order below $\SI{4}{K}$ which possesses a small FM component. 
Therefore, the synthesis of high-quality single crystals is essential for clarifying the intrinsic magnetic properties of CeGe$_{2-x}$. Moreover, the application of clean external tuning parameters such as magnetic field and pressure may tune the magnetic ground state as well as transport properties.

\setlength{\tabcolsep}{4.5pt}
\begin{table*}[hbt!]
	\centering
	\caption{Crystal structures, magnetic transition temperatures ($T_{\text{C}}$ and $T_{\text{N}}$), Sommerfeld coefficient ($\gamma$) and effective magnetic moment ($\mu_{\text{eff}}$) of CeGe$_{2-x}$ depending on different synthesis methods and Ge compositions. ``SC" represents single crystal in the ``Sample" column.}
	\label{tab:CeGe_comparison}
	\renewcommand\arraystretch{1.2}

	\begin{tabular}{ccccccccc}
		\hline
		\hline
		\boldmath$x$ & \textbf{Sample}& \textbf{Methods} & \textbf{Space group} & \boldmath$T_{\text{C}}$\text{ (K)} &\boldmath$T_{\text{N}}$\text{ (K)} & \boldmath$\gamma$ (\si{mJ/mol\cdot K^2}) & \boldmath$\mu_{\text{eff}}$ (${\mu_{\text{B}}/\text{Ce}}$) &\textbf{References} \\
		\hline
		0 & -- & -- & \textit{Imma} & 4.5 & -- & -- & -- & \cite{matthiasSuperconductivityFerromagnetismIsomorphous1958} \\
		0 & SC & -- & \textit{Imma} & 7 & -- & -- & 2.41 & \cite{yashimaThermalMagneticProperties1982} \\
		0 & -- & arc melting & \textit{Imma} & 6.9 & -- & -- & 2.45 & \cite{moriNewDenseKondo1985} \\
		 0 & -- & arc melting & \textit{Imma} & 4.3 & 7 & 120 & 2.38 & \cite{linThermalMagneticProperties2002} \\
		0.16 & -- & arc melting & \textit{Imma} & 4.3 & 7 & 113 & 2.37(3) & \cite{zanStudyMagneticOrdering2003} \\
		0.2 & SC & flux & \textit{C222}$_1$ & -- & 7.3 & 92 & 2.61(\textbf{\textit{H}$\perp$\textit{b}}), 2.77(\textbf{\textit{H}$\parallel$\textit{b}}) & This work \\
		0.24 & SC & flux & \textit{Imma} & 4 & 5, 7 & 110 & 2.53 & \cite{budkoPhysicalPropertiesCeGe2x2014,jayasekaraComplexMagneticOrdering2014} \\
		0.25 & SC & flux & \textit{C222}$_1$ & -- & 7 & -- & 2.25(3) & \cite{zhangSynthesisStructuralCharacterization2013} \\
		0.32 & -- & arc melting & \textit{I4$_{1}$/amd} & -- & 7 & 88 & 2.37(3) & \cite{zanStudyMagneticOrdering2003} \\
		0.34 & -- & arc melting & \textit{I4$_{1}$/amd} & 5.3 & 6.7 & 60 & 2.41 & \cite{nakanoElectricalResistivitySpecific2005} \\
		0.4 & SC & Czochralski & \textit{I4$_{1}$/amd} & 6 & -- & -- & -- & \cite{lambert-andronCrystalStructureProperties1990} \\
		0.46 & SC & Czochralski & \textit{I4$_{1}$/amd} & 4 & 4.5 & -- & -- & \cite{lambert-andronCoexistenceOrderedDisordered1994} \\
		\hline
		\hline
	\end{tabular}
\end{table*}
\setlength{\tabcolsep}{2.5pt}

In this work, we report the successful synthesis of high-quality single crystals of Ce$_4$Ge$_7$, which is confirmed to crystallize in the orthorhombic Nd$_4$Ge$_7$-type structure (space group \textit{C222}$_1$) and exhibits a Sommerfeld coefficient $\gamma $ of $\SI{92}{mJ/\text{mol-Ce}~K^2}$. A single AFM transition is observed at $T_{\text{N}} = \SI{7.3}{K}$.
Upon applying a magnetic field along the \textbf{\textit{b}}-axis, a metamagnetic transition occurs at $\mu_0H = \SI{1.3}{T}$ and $\SI{2}{K}$, which is accompanied by an anomalous Hall effect (AHE) dominated by the intrinsic Karplus–Luttinger (KL) mechanism.
With increasing pressure, $T_{\text{N}}$ first increases slightly and then decreases, exhibiting quantum critical behavior near $P_{\text{c}} = \SI{10.2}{GPa}$, as evidenced by a divergence of the $A$ coefficient, a maximum in the residual resistivity $\rho_0$, and non-Fermi liquid behavior, and then it possesses Fermi liquid ground state at higher pressures.

\section{\uppercase\expandafter{\romannumeral2}. EXPERIMENTAL METHODS}

Single crystals of Ce$_4$Ge$_7$ were grown using an indium flux \cite{zhangSynthesisStructuralCharacterization2013}. The Ce ingot, Ge granules, and In ingots were placed in an alumina crucible in a molar ratio of 1:1:40 and sealed in an evacuated quartz tube. The tube was heated up to $1150^\circ\text{C}$ and held at this temperature for 20 h, then it cooled down slowly to $500^\circ\text{C}$. Shiny cuboidal single crystals, with typical lengths of 2–5 mm, were obtained by centrifugation to remove the indium flux \cite{canfieldGrowthSingleCrystals1992a}. The chemical composition was determined via energy-dispersive x-ray spectroscopy (EDS) measurement on multiple areas, which provided an averaged atomic ratio of Ce : Ge = 1 : 1.80(2), and it is referred to as Ce$_4$Ge$_7$ through this paper. The crystal structure and orientation were determined via single-crystal X-ray diffraction (XRD) using a Bruker D8 Venture diffractometer with Mo $K_\alpha$ radiation at room temperature.

Magnetization measurements were performed using a Quantum Design (QD) Magnetic Property Measurement System (MPMS). The heat capacity measurement was performed in QD Physical Property Measurement System (PPMS). The electrical resistivity and Hall resistivity at ambient pressure were measured using PPMS. The electrical transport measurements under pressure were performed using diamond anvil pressure cell (DAC) in Teslatron-PT system equipped with an Oxford $^3$He refrigerator. Electrical contacts were made using highly conductive silver paste, and gold wires with a diameter of \SI{15}{\micro\meter} were employed. The sample used for electrical transport measurements under pressure was cut to approximate dimensions of \SI{15}{\micro\meter} $\times$ \SI{70}{\micro\meter} $\times$ \SI{120}{\micro\meter}. Resistance measurements were carried out with an excitation current of \SI{31.6}{\micro\ampere} flowing within the \textbf{\textit{ac}}-plane. Daphne oil 7373 served as the pressure transmitting medium to establish a quasi‑hydrostatic pressure environment. Pressure was calibrated by measuring the ruby fluorescence spectrum at room temperature \cite{grassetCalibrationRubyFluorescence2001a}.

  \begin{figure}
	\includegraphics[angle=0,width=0.5\textwidth]{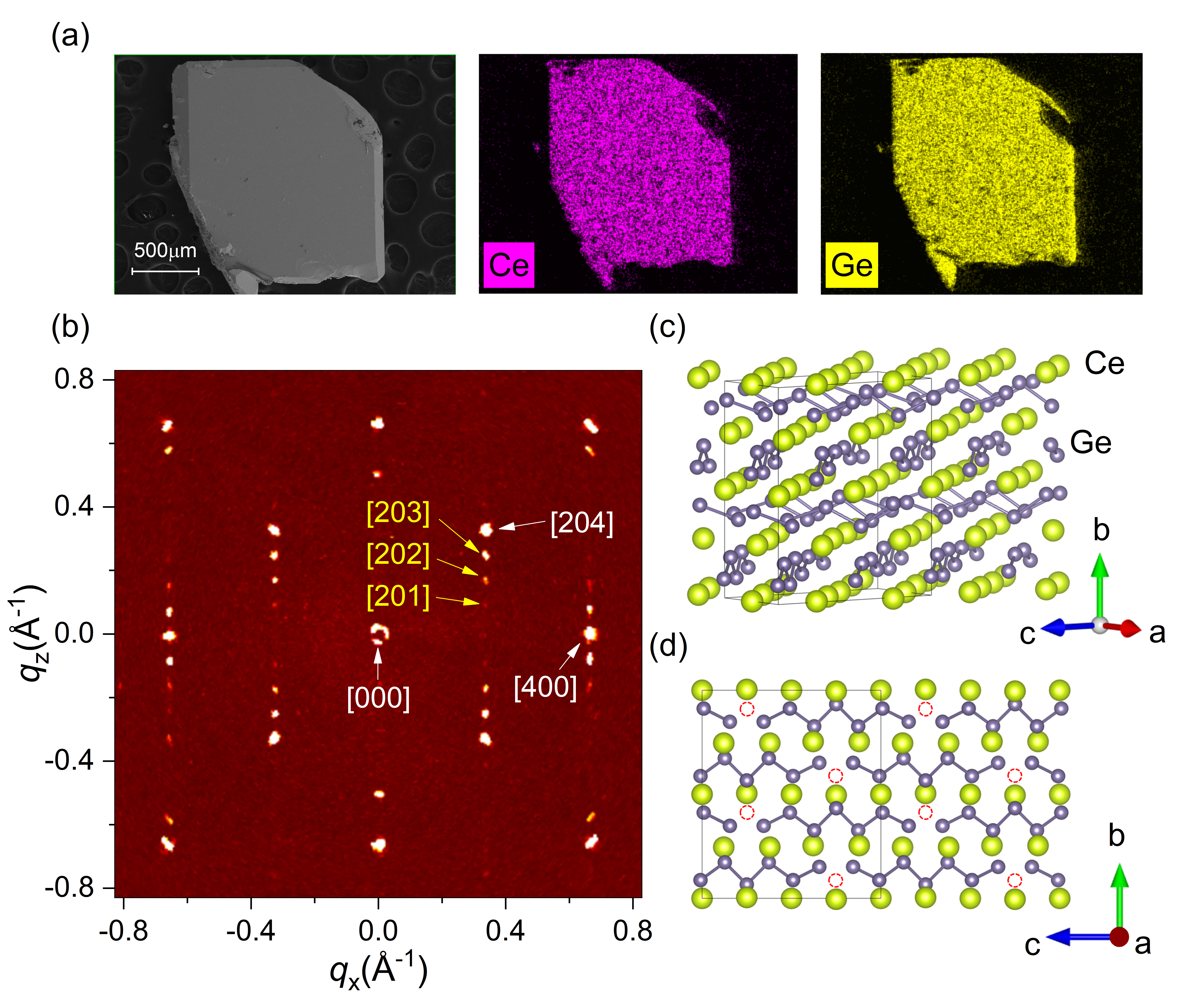}
	\vspace{-12pt} \caption{\label{Figure1}  (a) Scanning electron microscopy (SEM) image of a representative Ce$_4$Ge$_7$ single crystal and the corresponding elemental mapping. (b) X-ray diffraction pattern on the $h0l$ plane. Superstructure reflections relative to the parent \textit{Imma} structure are indicated by the yellow arrows.
		(c) Crystal structure (space group \textit{C222$_{1}$}) and (d) The view from [1 0 0] direction of Ce$_4$Ge$_7$. The red dotted circle in (d) indicates the Ge vacancies.}
	\vspace{-12pt}
 \end{figure}

\begin{table}[t]
	\caption{Crystal structure parameters for Ce$_4$Ge$_7$ determined from single-crystal XRD. This compound crystallizes in the orthorhombic space group \textit{C222}$_1$ (No. 20). The goodness of fit is 1.094, $R_1$ = 0.0529, $\omega R_2$ = 0.1195.}
	\label{table1}
	\centering
	\renewcommand\arraystretch{1.2}
	\begin{tabular}{ccccc}
		\firsthline
		\hline
		Lattice  &\textit{a}(\AA)&\textit{b}(\AA)&\textit{c}(\AA)&\textit{V}(\AA$^3$)\\
		parameters & 6.0445(6) & 14.0658(14) & 12.0707(13) & 1026.26(18) \\
		\firsthline
		\hline
		Atoms & \textit{x} & \textit{y} & \textit{z} &\textbf{$U$}\\
		 \hline

		 Ce1 & 0.2399(3) & -0.25387(12) & -0.37883(12) & 0.0070(5) \\
		 Ce2 & 0.50000 & -0.49580(17) & -0.25000 & 0.0073(6) \\
		 Ce3 & 0.0122(4) & -0.50000 & 0.00000 & 0.0086(6) \\
		 Ge1 & 0.2485(5) & -0.3269(3) & -0.1270(3) & 0.0145(8) \\
		 Ge2 & 0.00000 & -0.4332(4) & -0.25000 & 0.0148(10) \\
		 Ge3 & 0.5144(6) & -0.4122(2) & 0.0089(3) & 0.0180(8) \\
		 Ge4 & 0.3123(8) & -0.1514(3) & -0.1555(4) & 0.0386(13) \\
		\hline
		\hline
	\end{tabular}
\end{table}

\section{\uppercase\expandafter{\romannumeral3}. RESULTS AND DISCUSSION }

\subsection{A. Crystal Structure and Basic Physical Properties}

Figure \ref{Figure1}(a) shows the scanning electron microscopy (SEM) image of Ce$_4$Ge$_7$ single crystal. The corresponding elemental mapping demonstrates the homogeneity of the sample. The refinement of single-crystal XRD results indicates that Ce$_4$Ge$_7$ crystallizes in the orthorhombic Nd$_4$Ge$_7$-type structure with space group \textit{C222}$_1$, isostructural with Nd$_4$Ge$_7$ \cite{venturiniNewOrderedThSi2type1999}, Pr$_4$Ge$_7$ \cite{shcherbanCrystalStructureCompound2009}, and Sm$_4$Ge$_7$ \cite{zhangSynthesisStructuralCharacterization2013}. Table~\ref{table1} lists the refined structural parameters of Ce$_4$Ge$_7$, which contains three inequivalent Ce sites and four inequivalent Ge sites. The three inequivalent Ce sites all exhibit low local symmetry (site symmetries $C_1$ for Ce1 and $C_2$ for Ce2 and Ce3), which would account for the rich magnetism in CeGe$_{2-x}$ system \cite{matthiasSuperconductivityFerromagnetismIsomorphous1958,yashimaThermalMagneticProperties1982,moriNewDenseKondo1985,gokhaleCeGeCeriumGermaniumSystem1989,lambert-andronCrystalStructureProperties1990,schobinger-papamantellosStructuresMagneticProperties1991,lambert-andronCoexistenceOrderedDisordered1994,linThermalMagneticProperties2002,zanStudyMagneticOrdering2003,nakanoElectricalResistivitySpecific2005,zhangSynthesisStructuralCharacterization2013,budkoPhysicalPropertiesCeGe2x2014,jayasekaraComplexMagneticOrdering2014}. However, a detailed determination of the magnetic exchange interactions and the magnetic structure of Ce$_4$Ge$_7$ calls for further investigation by techniques such as neutron scattering.

Figure \ref{Figure1}(c) shows the crystal structure of Ce$_4$Ge$_7$. The Ge and Ce atomic layers stack alternately along the \textbf{\textit{b}}-axis. Each Ge layer consists of zigzag chains running along either the $[2\,0\,1]$ or $[-2\,0\,1]$ direction. Two adjacent layers of Ge zigzag chains which are separated by a Ce layer are oriented mutually perpendicular. In Ge zigzag chains, one Ge vacancy is formed per eight Ge atoms, and these vacancies themselves form chains along the \textbf{\textit{a}}-axis, as shown in Fig.~\ref{Figure1}(d). As a result, these periodically arranged Ge vacancies form a $\sqrt{2} \times 2\sqrt{2} \times 1$ superstructure, resulting in a unit cell four times larger than that of the parent \textit{Imma} structure. The superstructure reflections are also confirmed in the X-ray diffraction pattern on the $h0l$ plane [Fig.~\ref{Figure1}(b)], as indicated by the yellow arrows.

\begin{figure}[t]
	\includegraphics[angle=0,width=0.4\textwidth]{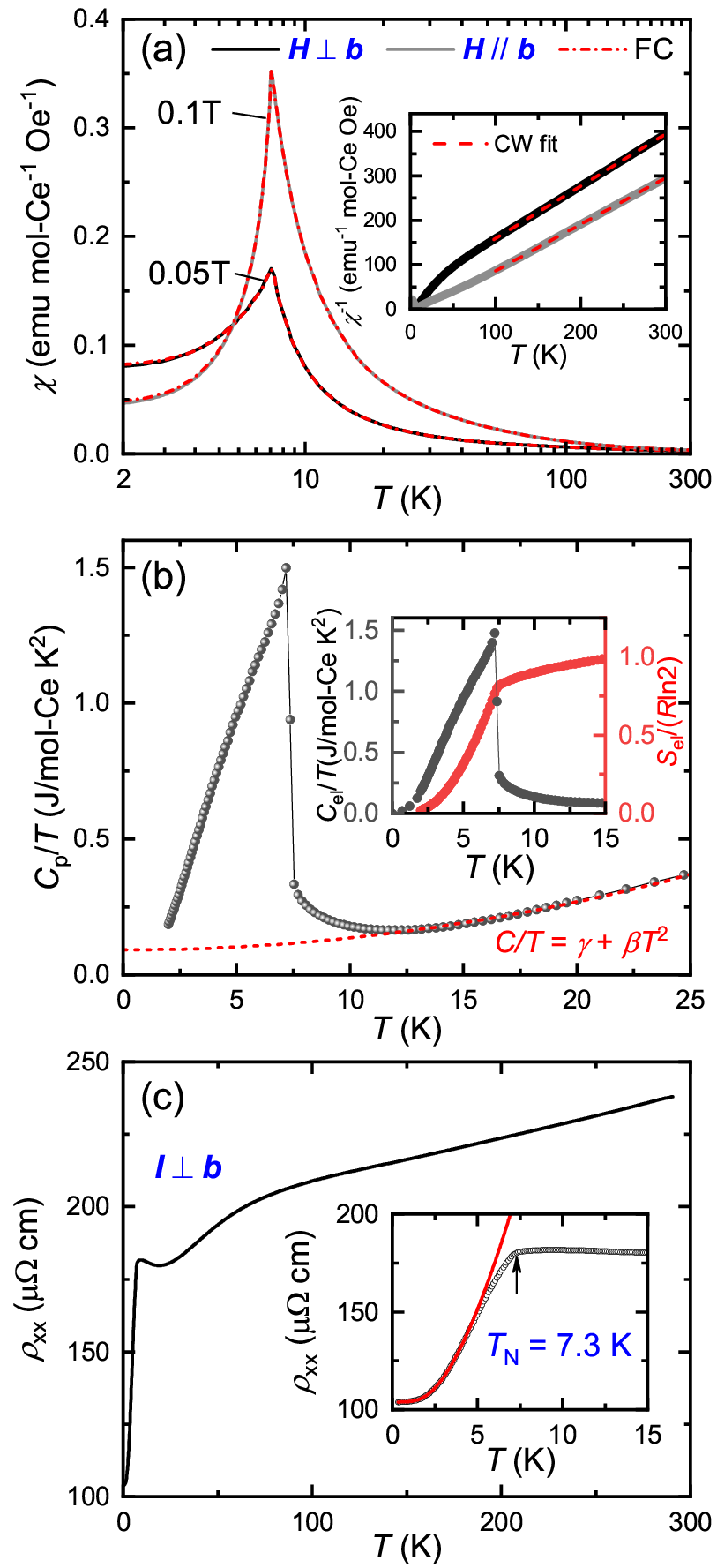}
	\vspace{-12pt} \caption{\label{Figure2} (a) Zero-field-cooled (ZFC) (solid lines) and field-cooled (FC) (dashed-dotted lines) magnetic susceptibility $\chi(T)$ for \textbf{\textit{H}~$\perp$~\textit{b}} and \textbf{\textit{H}~$\parallel$~\textit{b}}. The inset shows the inverse susceptibility $\chi^{-1}(T)$; the red dashed line represents a Curie–Weiss fit to the data in the temperature range from \SI{100}{K} to \SI{300}{K}.
		(b) Temperature dependence of the specific heat divided by temperature $C_p/T$. The red dashed line represents a fit to the function $C_p/T = \gamma + \beta T^2$. The inset shows the electronic specific heat $C_{\text{el}}/T$ (left axis) and the corresponding electronic entropy $S_{\text{el}}$ (right axis), where $R$ is ideal gas constant. 
		(c) Temperature dependence of the electrical resistivity $\rho(T)$. Inset: $\rho(T)$ below \SI{15}{K} with the fitting in red solid line as described in text.
		}
	\vspace{-12pt}
\end{figure}

\begin{figure*}[t]
	\includegraphics[angle=0,width=1 \textwidth]{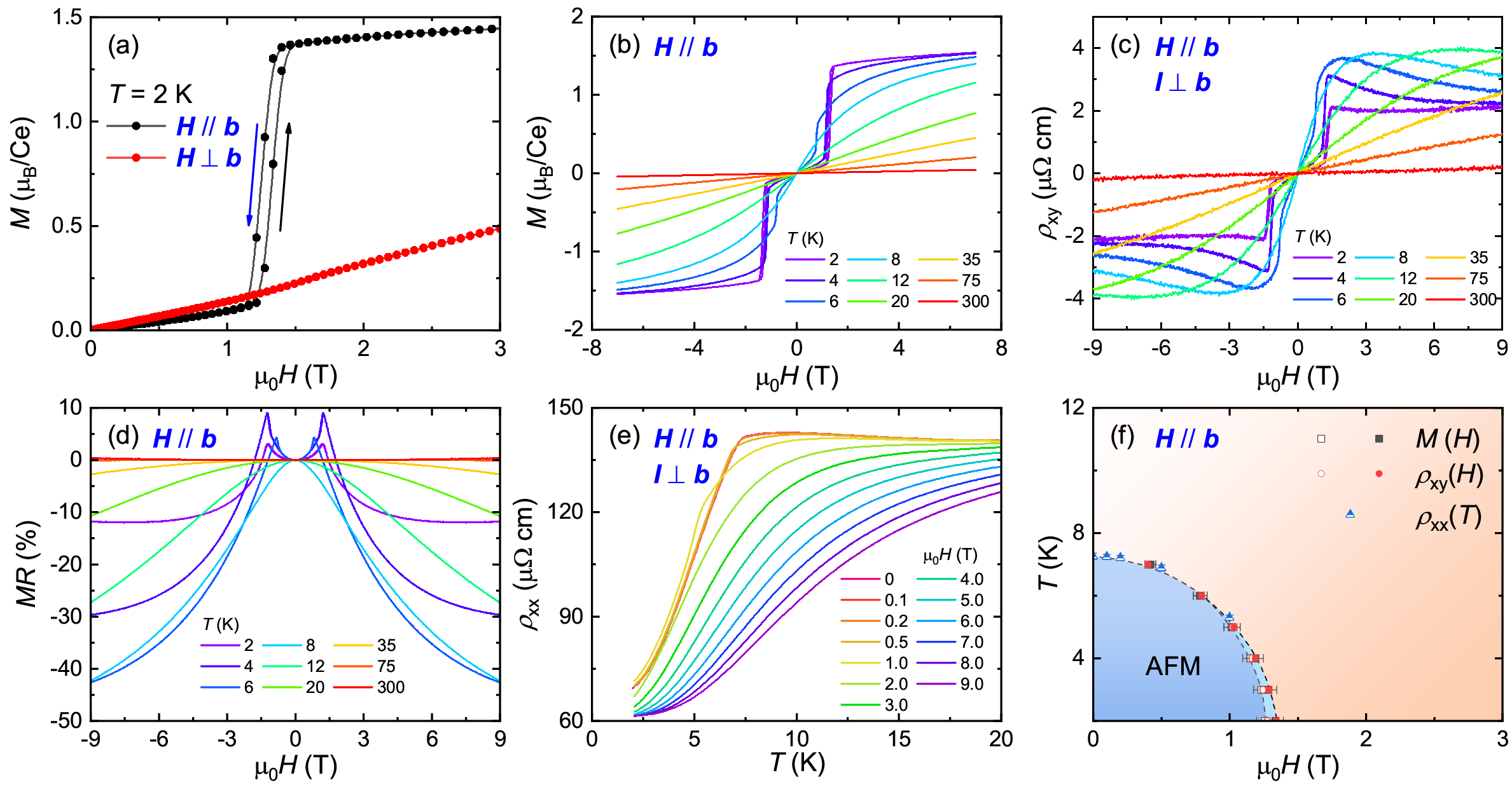}
	\vspace{-12pt} \caption{\label{Figure3} (a) Magnetization curves $M(H)$ of Ce$_4$Ge$_7$ at \SI{2}{K}. 
		(b) Isothermal magnetization curves at selected temperatures. 
		(c) Isothermal Hall resistivity $\rho_{xy}(H)$ curves. 
		(d) Isothermal magnetoresistance MR $= [\rho_{xx}(H) - \rho_{xx}(0)]/\rho_{xx}(0) \times 100\%$ curves. 
		(e) Temperature dependence of resistivity $\rho(T)$ at fixed magnetic fields. 
		(f) Magnetic field-temperature phase diagram of Ce$_4$Ge$_7$. Three individual single crystals were used for the magnetization, Hall resistivity, magnetoresistance and resistivity measurements, respectively, and the results are mutually consistent across the phase diagram. Solid and open symbols represent data measured upon increasing field and decreasing field, respectively. Representative temperature data are presented here for clarity, with the complete set provided in the Supplementary Information \cite{zotero-item-3966}.}
	\vspace{-12pt}
\end{figure*}

The temperature dependence of magnetic susceptibility $\chi(T)$ with \textbf{\textit{H}~$\perp$~\textit{b}} and \textbf{\textit{H}~$\parallel$~\textit{b}} is presented in Fig.~\ref{Figure2}(a). A single sharp cusp is observed at \SI{7.4}{K}, signaling the AFM transition. The overlapping ZFC and FC curves, in combination with the absence of hysteresis in the low-field $M$-$H$ curves [see Fig.~\ref{Figure3}(a)], indicate that there is not a detectable FM component  \cite{zhangSynthesisStructuralCharacterization2013}. 
The inverse susceptibility $\chi^{-1}(T)$ was fitted with the Curie-Weiss law, $\chi(T) = C / (T - \theta_{\text{CW}})$, as shown in the inset of Fig.~\ref{Figure2}(a), where $C$ is the Curie constant and $\theta_{\text{CW}}$ is the Curie-Weiss temperature. 
The fit yields $\theta_{\text{CW}} = \SI{-35}{K}$ for \textbf{\textit{H}~$\perp$~\textit{b}} and $\SI{17}{K}$ for \textbf{\textit{H}~$\parallel$~\textit{b}}, and an effective magnetic moment of $\mu_{\text{eff}} = \SI{2.61}{\mu_B/\text{Ce}}$ for \textbf{\textit{H}~$\perp$~\textit{b}} and $\SI{2.77}{\mu_B/\text{Ce}}$ for \textbf{\textit{H}~$\parallel$~\textit{b}}, respectively. These values are close to the theoretical value of $\SI{2.54}{\mu_B/\text{Ce}}$ for a free Ce$^{3+}$ ion.

\nocite{fertTheoryHallEffect1987a}

A clear jump in the specific heat $C_\text{p}$ is observed at \SI{7.3}{K} [Fig.~\ref{Figure2}(b)]. In the paramagnetic state, the $C_\text{p}/T$ data from \SIrange{12}{20}{K} were fitted using $C_\text{p}/T = \gamma + \beta T^2$, as shown in Fig.~\ref{Figure2}(b). The fit yields a Sommerfeld coefficient $\gamma = 92~\mathrm{mJ/\text{mol-Ce}~K^2}$, which is comparable to other CeGe$_{2-x}$ compounds. As listed in Table~\ref{tab:CeGe_comparison}, $\gamma$ generally decreases with increasing Ge deficiency, suggesting that Ge vacancies weaken the electron correlation strength in this system. The electronic specific heat $C_{\text{el}}/T$ was obtained by subtracting the fitted lattice contribution from the total specific heat. To estimate the entropy, $C_{\text{el}}/T$ below \SI{2}{K} was extrapolated to zero temperature. The electronic entropy $S_{\text{el}}$ is shown in the inset of Fig.~\ref{Figure2}(b). At $T_{\text{N}}$, the electronic entropy reaches only $0.82R\ln 2$, corresponding to an estimated Kondo temperature $T_{\text{K}}$ of \SI{2.6}{K} \cite{desgrangesSpecificHeatKondo1982}. $S_{\text{el}}$ recovers to $R\ln 2$ around \SI{15}{K}. This behavior indicates the presence of Kondo screening in Ce$_4$Ge$_7$, a phenomenon commonly observed in heavy-fermion compounds \cite{zhangStructuralMagneticProperties2020,zhangStructuralPhysicalProperties2024}. In the above analysis, we cannot estimate the contribution of crystalline electric field (CEF) effects to the specific heat, which calls for further neutrons experiments.

Figure \ref{Figure2}(c) shows the temperature dependence of the electrical resistivity $\rho(T)$ with the current applied within the \textbf{\textit{ac}}-plane. The $\rho(T)$ curve exhibits a broad hump around \SI{70}{K}, which is likely typical behavior in Ce-based compounds due to  the CEF excitations \cite{yeMagneticPropertiesLayered2022,zhangMagneticStatesKondo2026}.
At \SI{7.3}{K}, a clear kink appears in the $\rho(T)$ curve, corresponding to the AFM transition, consistent with the susceptibility and specific heat results. 
Below $T_{\text{N}}$, the resistivity drops rapidly, and the low-temperature resistivity can be fitted by the AFM spin-wave model \cite{fontesElectronmagnonInteraction$RmathrmNiBC$1999}, as shown in the inset of Fig.~\ref{Figure2}(c). The fitting formula is given by
\begin{equation} \label{eq2}
	\
	\rho(T) = \rho_0 + A T^2 + b \Delta^2 \sqrt{\frac{T}{\Delta}} \exp\!\left(-\frac{\Delta}{T}\right)\notag\\
\end{equation}

\begin{equation} \label{eq3}
	\
	\times \left[ 1 + \frac{2}{3}\frac{T}{\Delta} + \frac{2}{15}\left(\frac{T}{\Delta}\right)^2 \right],
	\
\end{equation}
where $\rho_0$ is the residual resistivity, the $A T^2$ term is the Fermi liquid contribution. The parameter $b$ represents the strength of the spin-wave scattering, while $\Delta$ corresponds to the spin-wave energy gap, which reflects the magnitude of the magnetocrystalline anisotropy. The fitted parameters are $\rho_0 = \SI{103.9(0)}{\micro\ohm\cm}$, $A = \SI{0.28(2)}{\micro\ohm\cm\per\K\squared}$, $b = \SI{2.60(2)}{\micro\ohm\cm\per\K\squared}$, and $\Delta = \SI{7.28(5)}{\K}$. However, the spin-wave gap and the magnon dispersion relation in Ce$_4$Ge$_7$ remain to be further explored experimentally.

\subsection{B. Metamagnetic Transition and Anomalous Hall Effect}

Figure \ref{Figure3}(a) shows the $M(H)$ curves of Ce$_4$Ge$_7$ measured at \SI{2}{K}, where pronounced magnetic anisotropy is observed. This is a common feature of the CeGe$_{2-x}$ series  \cite{lambert-andronCrystalStructureProperties1990,lambert-andronCoexistenceOrderedDisordered1994,budkoPhysicalPropertiesCeGe2x2014}. The $M(H)$ curve for \textbf{\textit{H} $\parallel$ \textit{b}} exhibits a sharp jump at \SI{1.3}{T}, indicating a metamagnetic transition. The presence of hysteresis between the field-increasing and field-decreasing branches suggests that the transition is first-order. Given that the magnetization rapidly saturates above this transition and the saturation moment $\SI{1.54}{\mu_B/\text{Ce}}$ ($\SI{7}{T}$, $\SI{2}{K}$) is close to $\SI{1.5}{\mu_B/\text{Ce}}$ which was reported for the FM phase in CeGe$_{2-x}$ \cite{lambert-andronCrystalStructureProperties1990}, we conclude that at \SI{1.3}{T}, Ce$_4$Ge$_7$ undergoes a transition from an AFM state to the spin-polarized state.  Besides, the reduction of saturation moment from the theoretical value of $\SI{2.14}{\mu_B/\text{Ce}}$ might originate from the Kondo screening effect and/or CEF effects. 

The metamagnetic transition being to the spin-polarized state is further corroborated by the Hall resistivity and magnetoresistance, as shown in Figs.~\ref{Figure3}(c) and (d). The Hall resistivity exhibits a pronounced AHE at the metamagnetic transition. The magnetic field at which this AHE signal appears corresponds well with the metamagnetic transition observed in the $M$-$H$ curves, as depicted in Fig.~\ref{Figure3}(b). Similarly, the isothermal magnetoresistance shows distinct kinks at the same critical field of the metamagnetic transition, and the negative magnetoresistance at higher fields is consistent with most ferromagnetic materials.

The critical field of this metamagnetic transition is gradually suppressed with increasing temperature, decreasing from \SI{1.3}{T} at \SI{2}{K} to \SI{0.4}{T} at \SI{7}{K}, as illustrated in Figs.~\ref{Figure3}(b)-(d). By combining temperature-dependent resistivity measurements at various magnetic fields [Fig.~\ref{Figure3}(e)], we constructed the magnetic field-temperature ($B$--$T$) phase diagram of Ce$_4$Ge$_7$, as shown in Fig.~\ref{Figure3}(f). Above $T_{\text{N}} = \SI{7.3}{K}$, no discernible transitions is observed in the Hall resistivity, magnetoresistance, and magnetization measurements. Instead, the paramagnetic state transforms to a spin-polarized state with increasing field in a crossover behavior.

 \begin{figure}[t]
 	\includegraphics[angle=0,width=0.45\textwidth]{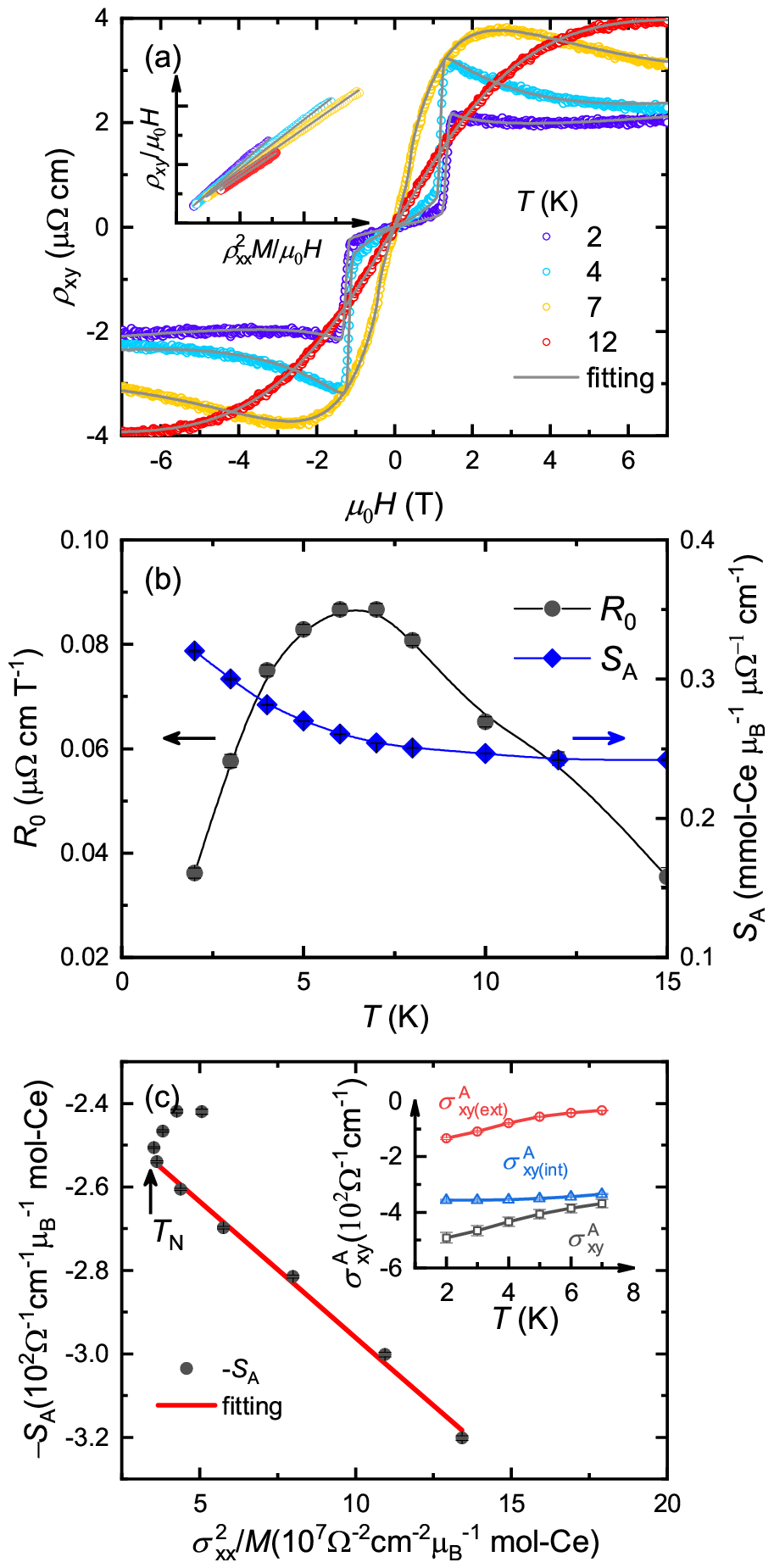}
 	\vspace{-12pt} \caption{\label{Figure4} (a) Hall resistivity $\rho_{xy}(H)$ at representative temperatures is shown here for clarity, with the complete set provided in the Supplementary Information \cite{zotero-item-3966}. The grey lines represent fitting results using the expression $\rho_{xy} = R_0 \mu_0 H + S_A \rho_{xx}^2 M$. Inset: the linear fitting of $\rho_{xy}/\mu_0 H$ as a function of $\rho_{xx}^2 M/\mu_0 H$. 
 		(b) Temperature dependence of the fitting parameters $R_0$ (left axis) and $S_A$ (right axis). (c) Linear fitting of $-S_{A}$ as a function of $\sigma^{2}_{xx}/M$ and the inset shows temperature dependence of the total anomalous Hall conductivity $\sigma_{xy}^{A}$ and its separated intrinsic ($\sigma_{xy,\text{int}}^{A}$) and extrinsic ($\sigma_{xy,\text{ext}}^{A}$) components.}
 	\vspace{-12pt}
 \end{figure}

To investigate the origin of the AHE in Ce$_4$Ge$_7$, we first separated the normal Hall contribution $\rho_{xy}^N$ from the total Hall resistivity $\rho_{xy}$ using \cite{leeHiddenConstantAnomalous2007}:
\begin{equation}\label{eq1}
\
\rho_{xy} = \rho_{xy}^N + \rho_{xy}^A = R_0 \mu_0 H + S_\text{A} \rho_{xx}^{2} M,
\
\end{equation}

\noindent where $\rho_{xy}^N = R_0 \mu_0 H$ is the normal Hall resistivity, and $\rho_{xy}^A=S_\text{A} \rho_{xx}^2 M$ represents the contribution from AHE.  
In order to determine the parameters $R_0$ and $S_\text{A}$, $\rho_{xy}\mu_0 H$ curves above the critical field ($\mu_0 H_c$) of the metamagnetic transition was fitted using $\rho_{xy}/(\mu_0 H)$ versus $\rho_{xx}^2 M/(\mu_0 H)$, where $R_0$ and $S_\text{A}$ represent the intercept and slope obtained from the fitting, respectively. Excellent linear relationships are observed, as shown in the inset of Fig.~\ref{Figure4}(a). The obtained $R_0$ and $S_\text{A}$ at various temperatures are exhibited in Fig.~\ref{Figure4}(b). 
The fitting parameter $R_0$ reflects the carrier concentration via the relation $R_0 = -1/(ne)$. In the temperature range of \SIrange{2}{15}{K}, $R_0$ remains positive, indicating that holes are the dominant charge carriers in Ce$_4$Ge$_7$, with a carrier concentration of the order of \SI{1e22}{cm^{-3}}. Furthermore, the non-monotonic temperature dependence of $R_0$, which peaks near $T_{\text{N}}$, could be attributed to either the reconstruction of the Fermi surface below the antiferromagnetic transition or modifications of the scattering mechanisms \cite{hamzicHallEffectHeavyfermion1988,siegristMagnetotransportHeavyfermionCompound1986}.
It's noteworthy that $S_\text{A}$ decreases monotonically with increasing temperature, deviating from the case that KL mechanism dominate the AHE, where $S_A$ should be independent of temperature \cite{wangAnomalousHallEffect2016}. This indicates the non-negligible contribution of extrinsic skew scattering and/or side jump mechanisms.
Overall, using the obtained parameters, $\rho_{xy}$ over the entire magnetic field range can be well fitted, as shown in Fig.~\ref{Figure4}(a), which also  precludes the presence of topological Hall effect in Ce$_4$Ge$_7$ \cite{kanazawaLargeTopologicalHall2011}.

To further separate the intrinsic and extrinsic contributions to the AHE, we use the Tian-Ye-Jin model \cite{tianProperScalingAnomalous2009,yeTemperatureDependenceIntrinsic2012}:
\begin{equation}\label{eq3.3}
	\sigma^{A}_{xy} = -(a_{\text{sk}} \sigma^{-1}_{xx0} + b_{\text{sj}} \sigma^{-2}_{xx0}) \sigma^{2}_{xx} + \sigma^{A}_{xy,\text{int}},
\end{equation}
where $\sigma^{A}_{xy}$ is the total anomalous Hall conductivity, $a_{\text{sk}}$ and $b_{\text{sj}}$ are constants representing the contribution of extrinsic skew scattering and side jump, respectively, $\sigma^{-1}_{xx0}$ is the residual conductivity, and $\sigma^{A}_{xy,\text{int}}$ is the intrinsic anomalous Hall conductivity. $\sigma^{A}_{xy}$ can be obtained from the relation $\sigma^{A}_{xy} \approx -\rho^{A}_{xy} / \rho^{2}_{xx}$ (since $\rho^{2}_{xx} \gg \rho^{2}_{xy}$), with $\rho^{A}_{xy} = S_A \rho^{2}_{xx} M$. Considering $\sigma^{A}_{xy,\text{int}}$ is proportional only to the magnetization $M$ and independent of scattering from phonons or impurities \cite{zengLinearMagnetizationDependence2006,leeHiddenConstantAnomalous2007,wuLinearMagnetizationDependence2021}, it can be written as $\sigma^{A}_{xy,\text{int}} = S_{A,\text{int}} M$. Dividing both sides of Eq.~\ref{eq3.3} by $M$, we obtain:
\begin{equation}\label{eq4}
	-S_A = c_{\text{ext}} \frac{\sigma^{2}_{xx}}{M} + S_{A,\text{int}},
\end{equation}
where $c_{\text{ext}} = -(a_{\text{sk}} \sigma^{-1}_{xx0} + b_{\text{sj}} \sigma^{-2}_{xx0})$ represents the contribution from extrinsic mechanisms. Using Eq.~\ref{eq4}, we performed a linear fit of $-S_A$ versus $\sigma^{2}_{xx}/M$ below $T_{\text{N}}$, as shown in Fig.~\ref{Figure4}(c). The fit yields an intercept $S_{A,\text{int}} = -231(2) \ \Omega^{-1}\text{cm}^{-1} \mu_{\text{B}}^{-1} \text{mol-Ce}$ and a slope $c_{\text{ext}} = -0.65(2) \ \mu\Omega \cdot \text{cm}$. The temperature dependence of the total anomalous Hall conductivity $\sigma_{xy}^{A}$ and its separated intrinsic ($\sigma_{xy,\text{int}}^{A}$) and extrinsic ($\sigma_{xy,\text{ext}}^{A}$) components are shown in the inset of Fig.~\ref{Figure4}(c). It could be concluded that the intrinsic KL mechanism dominates the AHE and exhibits a relatively weak temperature dependence, whereas the extrinsic contribution becomes increasingly significant upon cooling. 
The extrinsic skew scattering ($\rho_{xy,\text{sk}}^A \propto \rho_{xx0} M$) likely plays a subordinate role. Because the electrical conductivity of Ce$_4$Ge$_7$ at \SI{2}{K} is approximately $10^4$ $\Omega^{-1}$ cm$^{-1}$, which places it in the good metal regime ($10^4$–$10^6$ $\Omega^{-1}$ cm$^{-1}$), whereas skew scattering was usually considered to dominate in the high conductivity regime ($\sigma_{xx} > 10^6$ $\Omega^{-1}$ cm$^{-1}$) \cite{onodaIntrinsicExtrinsicAnomalous2006,nagaosaAnomalousHallEffect2010}.

\begin{figure}[t]
	\includegraphics[angle=0,width=0.38\textwidth]{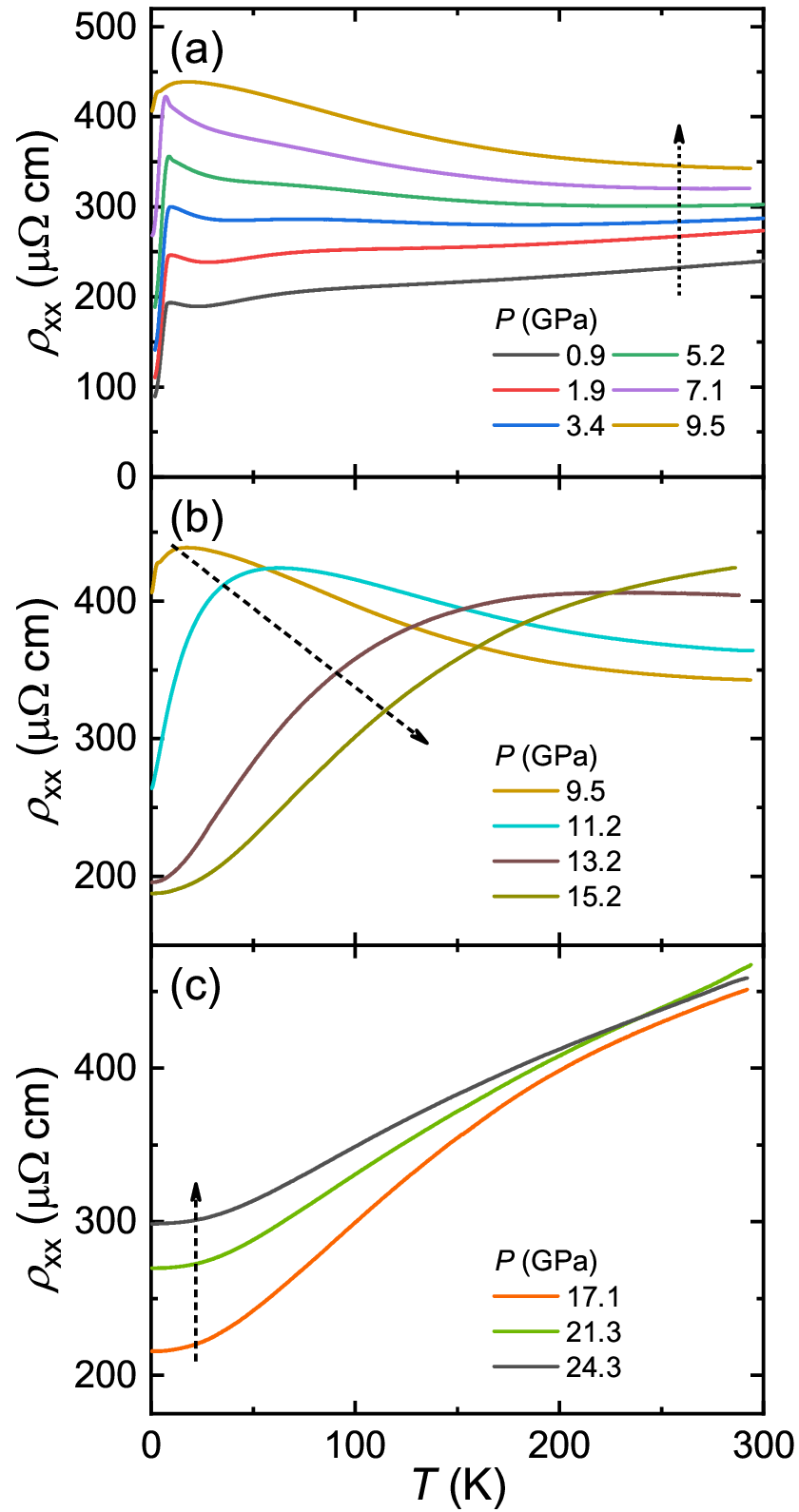}
	\vspace{-12pt} \caption{\label{Figure5} Temperature dependence of the electrical resistivity $\rho_{\text{xx}}(T)$ (\textbf{\textit{I}}~$\perp$~\textbf{\textit{b}}) of sample 1 (S1) (a) from \SIrange{0.9}{9.5}{GPa}, (b) from \SIrange{9.5}{15.2}{GPa}, (c) from \SIrange{17.1}{24.3}{GPa}. The black arrow indicates the trend of the evolution of $\rho_{\text{xx}}(T)$ curves as the pressure increases.}
	\vspace{-12pt}
\end{figure}

\begin{figure}[t]
	\includegraphics[angle=0,width=0.5\textwidth]{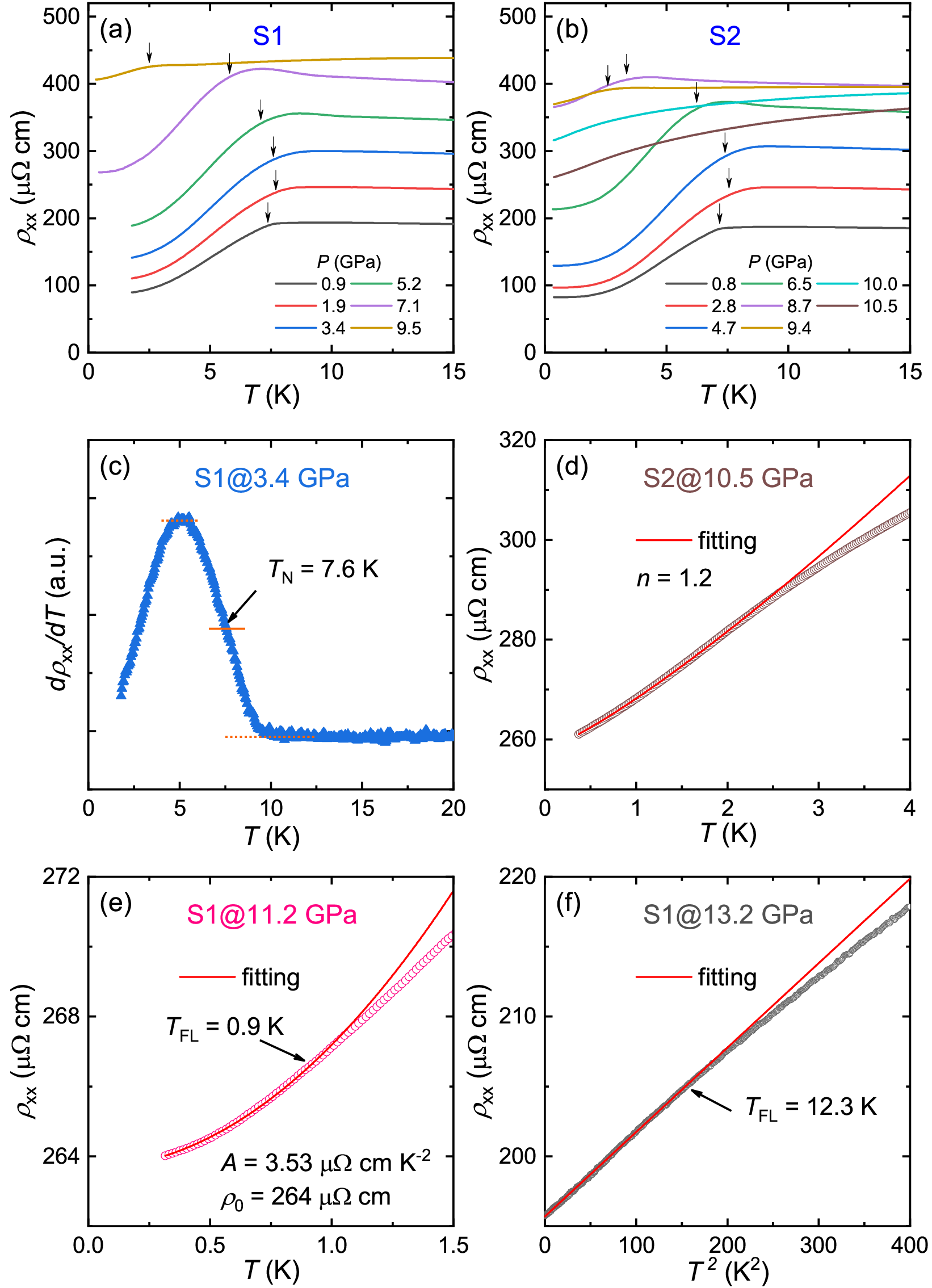}
	\vspace{-12pt} \caption{\label{Figure6} The $\rho_{\text{xx}}(T)$ curves under pressure for samples (a) S1 and (b) S2 below $\SI{15}{K}$, where the black arrows indicate the AFM transition temperature $T_{\text{N}}$. (c) The determination of $T_{\text{N}}$ in $d\rho_{\text{xx}}/dT$ curve at a representative pressure of  \SI{3.4}{GPa} for S1, as marked by a black arrow. (d) $\rho_{\text{xx}}(T)$ curves for S2 at \SI{10.5}{GPa} below \SI{4}{K}, along with the fitting results using $\rho_{\text{xx}}(T) = \rho_0 + A T^n$ for the non-Fermi liquid behavior. (e) $\rho_{\text{xx}}(T)$ curves for S1 at \SI{11.2}{GPa} below \SI{1.5}{K}, along with the fitting results using $\rho_{\text{xx}}(T) = \rho_0 + A T^2$ for the Fermi liquid behavior. (f) $\rho_{\text{xx}}(T)$ versus $T^2$ curve and Fermi liquid fitting at a representative pressure of  \SI{13.2}{GPa} for S1. The Fermi liquid temperature $T_{\text{FL}}$ is defined as the temperature above which the $\rho_{\text{xx}}(T^2)$ curve begins to deviate from linearity, as marked by a black arrow. Fitting results at other pressures are available in the Supplementary Information \cite{zotero-item-3966}.}
	\vspace{-12pt}
\end{figure}
 
\begin{figure}[t]
	\includegraphics[angle=0,width=0.5\textwidth]{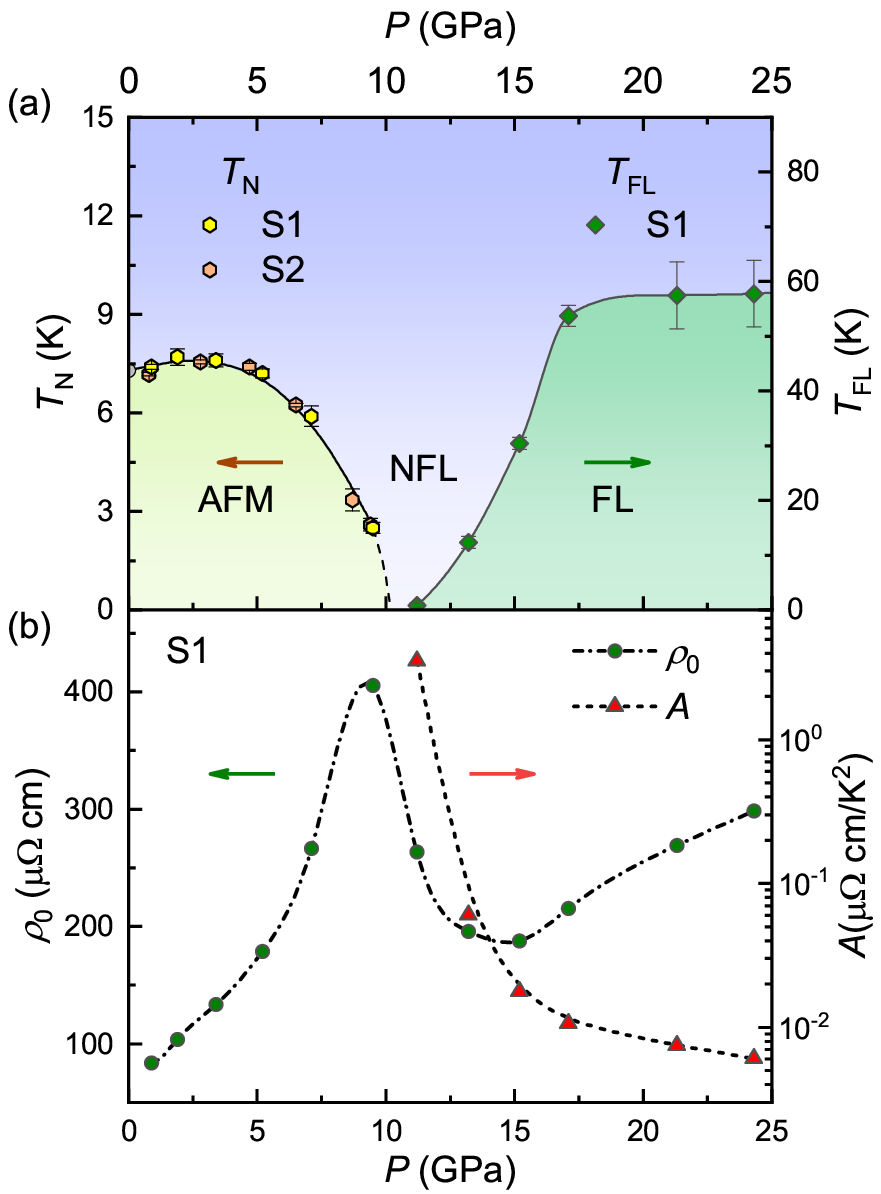}
	\vspace{-12pt} \caption{\label{Figure7} (a) Temperature-pressure ($T$-$P$) phase diagram for Ce$_4$Ge$_7$. The left and right axes correspond to the N\'{e}el temperature $T_{\text{N}}$ and the Fermi liquid temperature $T_{\text{FL}}$, respectively. The dashed line indicates the extrapolation of $T_{\text{N}}$ towards \SI{0}{K}, at $P_{\text{c}}=$ \SI{10.2}{GPa}. (b) The evolution of residual resistivity $\rho_0$ (left axis) and $A$ coefficient (right axis) under pressure.}
	\vspace{-12pt}
\end{figure}

\subsection{C. Pressure-Induced Quantum Critical Behavior}

To investigate the effect of pressure on the transport properties of Ce$_4$Ge$_7$, we performed electrical resistivity measurements under pressure using a diamond anvil cell (DAC). The overall resistivity of Ce$_4$Ge$_7$ increases with pressure from \SIrange{0.9}{9.5}{GPa} [Fig.~\ref{Figure5}(a)], while $T_{\text{N}}$ exhibits a non-monotonic evolution. As marked by black arrows in Fig.~\ref{Figure6}(a), $T_{\text{N}}$ of sample 1 (S1) first increases from \SI{7.3}{K} at ambient pressure to a maximum of \SI{7.7}{K} at \SI{1.9}{GPa}. Then, it is suppressed to \SI{2.5}{K} at \SI{9.5}{GPa} and becomes indiscernible at higher pressures. Similar results of sample 2 (S2) demonstrate good reproducibility, as shown in Fig.~\ref{Figure6}(b).

Above \SI{9.5}{GPa}, a broad hump emerges in $\rho_{\text{xx}}(T)$ [Fig.~\ref{Figure5}(b)], likely originating from Kondo scattering, as is common in many heavy-fermion systems \cite{stewartHeavyfermionSystems1984}.
As indicated by the dashed black arrow in Fig.~\ref{Figure5}(b), this hump shifts to higher temperatures with increasing pressure, indicating that the Kondo temperature $T_{\text{K}}$ increases with pressure. 
To conduct a quantitative study on the evolution of low-temperature resistance behavior with pressure, we fitted the $\rho(T)$ curves using the Fermi liquid expression $\rho(T) = \rho_0 + A T^2$ [as shown in Figs.~\ref{Figure6}(e) and (f)], where $\rho_0$ is the residual resistivity and the $A$ coefficient reflects the effective electron mass [$A \propto (m^*)^2$] \cite{terhaar90TheoryFermi1965}.
At pressures of \SI{11.2}{GPa} and above, the low-temperature resistivity exhibits typical Fermi liquid behavior. Fitting results indicate that the $A$ coefficient decreases rapidly with increasing pressure, changing by nearly three orders of magnitude from \SI{11.2}{GPa} to \SI{24.3}{GPa} [Fig.~\ref{Figure7}(b)]. This suggests a substantial reduction in the effective mass under pressure, possibly signaling an entry into a mixed-valence state \cite{watanabeQuantumCriticalPhenomena2023}.
Therefore, the dominance of the Kondo effect over the RKKY interaction leads to the ground state of Ce$_4$Ge$_7$ evolving from an AFM state to a Fermi liquid state with pressure. 
\setlength{\tabcolsep}{6pt}
\begin{table}[hbt!]
	\centering
	\caption{Comparison of the $A$ coefficient and the residual resistivity $\rho_0$ for various heavy-fermion compounds near their respective pressure-induced quantum critical points.}
	\label{tab:A_comparison}
	\renewcommand\arraystretch{1.2}
	\begin{tabular}{>{\color{black}}l>{\color{black}}c>{\color{black}}c>{\color{black}}c}
		\hline
		\hline
		Compound & $A$(\si{\micro\ohm\cm\per\K\squared}) & $\rho_0$(\si{\micro\ohm\cm}) & Reference \\
		\hline
		CeRh$_2$Si$_2$ & 0.012 & 0.8 & \cite{arakiPressureinducedSuperconductivityAntiferromagnet2002a} \\
		CeAl$_2$ & 0.35 & 15 & \cite{miyagawaElectronicStatesSingle2008a} \\
		CeIn$_3$ & 0.5 & 2 & \cite{knebelElectronicPropertiesCeIn2001} \\
		CeRh$_6$Ge$_4$ & 1.1 & 1.7 & \cite{shenStrangemetalBehaviourPure2020} \\
		CeCoGe$_{2.1}$Si$_{0.9}$ & 1.7 & 64 & \cite{alzamoraAntiferromagneticQuantumCriticality2007} \\
		Ce$_4$Ge$_7$ & 3.5 & $\sim$400 & This work \\
		CeRhIn$_5$ & 6$\text{(15T)}$ & 16 & \cite{knebelQuantumCriticalPoint2008} \\
		CePt$_2$In$_7$ & 8 & 10 & \cite{sidorovPressurePhaseDiagram2013} \\
		CeNiGe$_3$ & 9 & $\sim$200 & \cite{kotegawaPressureinducedSuperconductivityCeNiGe32006} \\
		Ce$_2$Ni$_3$Ge$_5$ & 11 & 8 & \cite{nakashimaChangeElectronicState2005} \\
		Ce$_2$NiGa$_{12}$ & 22 & 75 & \cite{kawamuraHighPressureProperties2014} \\
		CePdAl & 39 & 55 & \cite{zhaoQuantumcriticalPhaseFrustrated2019} \\
		\hline
		\hline
	\end{tabular}
\end{table}

Figure \ref{Figure7} summarizes the temperature--pressure ($T$--$P$) phase diagram of Ce$_4$Ge$_7$, together with the evolution of the residual resistivity $\rho_0$ and the $A$ coefficient. The behavior of $T_{\text{N}}$ under pressure is reminiscent of the Doniach phase diagram \cite{doniachKondoLatticeWeak1977}. The slight low pressure increase of $T_{\text{N}}$ indicates that the 4$f$ electrons in Ce$_4$Ge$_7$ are well localized at ambient pressure. The extrapolation of $T_{\text{N}}$ to \SI{0}{K} occurs near $P_{\text{c}} = \SI{10.2}{GPa}$. At $P_{\text{c}}$, the residual resistivity $\rho_0$ reaches its maximum, which likely related to enhanced AFM spin fluctuations. In addition, the $A$ coefficient shows a divergent behavior near $P_{\text{c}}$, indicating a diverging effective mass. These signatures are reminiscent of those observed in many pressure-induced quantum critical heavy-fermion systems. A comparison of the $A$ coefficient and the residual resistivity near the quantum critical point for representative compounds is summarized in Table~\ref{tab:A_comparison}. It should be noted that the $\rho(T)$ curves around $P_{\text{c}}$ exhibit non-Fermi liquid behavior, with an exponent $n$ less than 2, as shown in Fig.~\ref{Figure6}(d) for \SI{10.2}{GPa} and in the Supplementary Information \cite{zotero-item-3966} for \SI{10.0}{GPa}. Above $P_{\text{c}}$, the Fermi liquid temperature $T_{\text{FL}}$ increases monotonously with pressure. Therefore, a possible AFM QCP exists in Ce$_4$Ge$_7$ around $P_{\text{c}} = \SI{10.2}{GPa}$. While non-Fermi-liquid behavior and divergent effective masses are commonly observed around the AFM QCP in heavy-fermion systems, the microscopic origin of such QCPs remains a central challenge in strongly correlated quantum many-body theory. Key questions include whether they arise from itinerant spin-density-wave fluctuations or from local Kondo destruction, as well as the conditions and mechanisms for unconventional superconductivity. For Ce$_4$Ge$_7$, determining the nature of the AFM QCP will require further pressure-dependent studies, including quantum oscillations, Hall effect, and thermoelectric transport measurements. Intriguingly, unlike CeRhIn$_5$ and other systems where superconductivity masks the QCP, Ce$_4$Ge$_7$ exhibits no superconductivity down to \SI{0.3}{K}, thus providing a clean platform for exploring AFM quantum critical phenomena.

\section{\uppercase\expandafter{\romannumeral4}. SUMMARY}

In summary, we successfully synthesized high quality single crystals of Ce$_4$Ge$_7$ with orthorhombic Nd$_4$Ge$_7$-type structure (space group \textit{C222}$_1$). Measurements of magnetic susceptibility, specific heat and electrical resistivity demonstrate that it is a moderately correlated AFM material with $T_{\text{N}} = \SI{7.3}{K}$. Both magnetic field and pressure can effectively tune its magnetism. At \SI{2}{K}, the AFM state evolves into a spin-polarized state above a critical field of $\SI{1.3}{T}$, accompanied by an AHE. The analysis of $\rho_{xy}(T,B)$ indicates that the AHE in Ce$_4$Ge$_7$ mainly arises from the intrinsic mechanism. Under pressure, $T_{\text{N}}$ is first slightly enhanced and then suppressed. The behavior of the resistivity, together with the evolution of the residual resistivity $\rho_0$ and the $A$ coefficient, suggests the presence of an AFM QCP near $P_{\text{c}} = \SI{10.2}{GPa}$.
To further explore whether chemical doping, pressure, or reducing dimensionality can induce superconductivity or FM QCP in CeGe$_{2-x}$ with different stoichiometry is of high interest.

\section{ACKNOWLEDGMENTS}

 This work was supported by the National Key R\&D Program of China (No. 2022YFA1402200 and No. 2023YFA1406303), the National Natural Science Foundation of China (No. 12674175, No. 12550401, No. W2511006, No. U23A20580, No. 12350710785, and No. 12204159), the Zhejiang Provincial Natural Science Foundation of China (No. LRG26A040001).
\bibliography{CeGe1.8}

\end{document}